\documentstyle[12pt,aasms4]{article}
\def\be{\begin{equation}}
\def\en{\end{equation}}
 
\begin{document}
\title{Gravitational waves from galaxy clusters: A new observable effect}
\author{Vicent Quilis
, Jos\'e M$^{\underline{\mbox{a}}}$. Ib\'a\~nez
\and Diego S\'aez}
\affil{Departament d'Astronomia i
Astrof\'{\i}sica, Universitat de Val\`encia,
E-46100 Burjassot, Val\`encia, Spain}
\authoremail{vicente.quilis@uv.es}

\begin{abstract}
A rich galaxy cluster showing strong resemblance with the observed ones
is simulated. Cold dark matter spectrum, Gaussian statistics, flat universe, 
and two components -- baryonic gas plus dark matter particles -- are 
considered. We have calculated the gravitational-wave output during the epoch
of the fully nonlinear and nonsymmetric cluster evolution.
The amplitudes and frequencies of the resulting gravitational waves are 
estimated. Since frequencies are very small --of the order of
$10^{-17} \ Hz$ -- a complete 
pulse cannot be observed during an admissible integration time; 
nevertheless, it 
is proved that these waves can produce  an interesting secular effect 
which appears to be observable with current technology. 
 
\end{abstract} 


\section{Introduction}

Any dynamical and nonspherical self-gravitating astrophysical
-- or cosmological -- system of mass M and size R is a potential source of 
gravitational waves. The most powerful ones are those combining a very 
asymmetric shape together with a violent nonstationary evolution.

In present paper we are interested in the gravitational waves radiated by 
cosmological objects.
Flattened structures as superclusters and walls show great departures from 
spherical symmetry, but they are evolving in the mildly nonlinear regime 
and, consequently, their evolution is slow. Gravitational waves from these 
structures will be considered elsewhere.
Galaxy clusters evolve in the strongly nonlinear regime and their dynamics is 
expected to be faster than in the case of larger flattened structures; 
however, departures from spherical symmetry seem to be small (even if their 
corresponding initial conditions were fully asymmetric). In this case, high
density contrasts -- of the order of $10^{3}$ -- are reached and, consequently, 
fully nonlinear numerical simulations are necessary. This paper concerns 
with the characteristics of the gravitational radiation released during
galaxy cluster evolution.  

As it is well known, clusters contain dark matter ($\sim 90 \% $) and 
a subdominant baryonic component ($\sim 10 \% $). Since the dynamics and 
spatial distribution of these components are different, 
the ratio between the gravitational luminosities coming from 
these two components is not expected to be constant.
Numerical simulations 
including a baryonic component -- gravitationally coupled to the dark one --
are necessary to describe the evolution of that ratio. 

Taking into account the above discussion, a two components cluster, mimicking 
the observed ones, has been simulated numerically, paying particular attention
to estimate its gravitational luminosity and the features of the radiated 
gravitational waves. To this aim we have used the approach of considering 
that the sources satisfy the constraints of the so-called nearly Newtonian 
slow motion. As far as we know, similar calculations have not been performed 
up to now, probably due to some pessimistic point of view justified by the
combination of great numerical difficulties (3D calculations) together
with very disappointing small values expected for the
gravitational-wave luminosity, the amplitude, et cetera. 
However, our results have been more 
encouraging than we could foresee. A new observable effect has
been found.

Hereafter, $t$ stands for the cosmological time, 
$t_0$ is the age of the Universe. 
$\dot{X}$ stands for the derivative of the function X with respect to the 
cosmological time. 
The present value of the Hubble constant is  assumed to be
$50 \ Km \ s^{-1} \ Mpc^{-1}$.
Quantities $c$ and $G$  are
the speed of light and the gravitational constant, respectively.
Greek (Latin) indices run from 1 to 4 (1 to 3). 

\section{Cluster Model}

A flat universe, cold dark matter (CDM), and adiabatic energy 
density fluctuations are assumed. The corresponding spectrum is normalized by 
the condition $\sigma_{8}=0.63$. Constrained Gaussian realizations of the 
density field -- in the position space -- can be obtained by using the method 
proposed by Hoffman and Ribak (1991) and improved by Weygaert and Bertschinger 
(1996). We have used this powerful tool to obtain an initial density field 
-- at redshift z=100 -- containing the cluster seed centered in a box. The 
constraints have been introduced in such a way that, after evolution, our 
initial overdensity leads to a feasible rich cluster (see Quilis et al.
1998 for details). It is assumed that the 90 \% of the matter is 
CDM and the remaining, the baryonic one, is assumed to be a monoatomic ideal 
gas. The initial values for the velocity field of both components are 
identical. No assumptions about the symmetry of the object have been done
at all. The resulting departures from spherical symmetry correspond to an 
arbitrary statistical realization of the density field.
 
The cosmological hydrodynamic equations describing the
evolution of the baryonic component (see Peebles, 1980) 
are solved using a hydro-code
based on {\it modern high-resolution shock-capturing techniques}. 
This code was
described in Quilis et al. (1996). The motion of dark matter particles
is studied 
by means of a standard {\it Particle Mesh} code (Hockney and 
Eastwood 1988). The total gravitational field is computed by solving
the Poisson equation --which couples baryons and dark matter--
with a multidimensional method based on 
Fast Fourier Transform (Press et al. 1987). Details about our galaxy cluster
simulations are described in Quilis et al. (1998).  

At present time, the simulated cluster has: (i) a X-ray luminosity of 
$\sim 10^{44} \ erg/s$, (ii) a temperature of $\sim 3 \times 10^{7} \ K$, 
and a total mass inside the Abell radius ($3\ Mpc$) of 
$5.4 \times 10^{14} \ M_{\odot}$; hence, we are considering a rich cluster 
having features compatible with present observations (B\"{o}eringer, 1991; 
Peebles, 1994).

\section{Gravitational Radiation}

Our cluster evolves in a flat universe and it 
is observed with a detector of gravitational waves moving with the
cosmological expanding background. 
This is the most 
natural reference frame in Cosmology and it is locally Minkowskian
at any time.
The metric distance from the detector to
the cluster, $D(t)$, is proportional to the scale factor 
$a(t) \propto (1+z)^{-1}$. Its present value is assumed
to be $D(t_{0})=100 \ Mpc$.
Gravitons reaching the detector at present time were emitted by the cluster
at time $t_{_{E}}$ (emission event E).
Could we perform our calculations in
the Minkowskian space tangent 
to the Robertson-Walker spacetime at E?
The particle point of view is appropriate to answer this question.
The emission and propagation
of gravitons in the real and tangent spaces are now compared:
(i) Since graviton
emission only depends on the internal dynamics of the system,
and it is the same in both cases, the same
amount of gravitons with the same energies are emitted,
(ii) The background universe as well as the Minkowski empty spacetime 
are transparent to graviton propagation and,
(iii) we have studied the
null geodesics in both spacetimes to calculate the times $t_{_{RW}}$
(Robertson-Walker) 
and $t_{_{M}}$ (Minkowski) elapsed by a graviton in 
reaching detectors located at the 
same distances from the cluster. 
The resulting values of  $t_{_{RW}}$ and $t_{_{M}}$ appear to be identical
to first order in the parameter
$D(t)/6000$. This parameter is much smaller than unity because
the distance $D(t) \leq 100 \ Mpc$  is much
smaller than the horizon scale of $6000 \ Mpc$. 
This means that, in both cases, the detectors are receiving signals
coming from the same retarded positions of the cluster.
After these considerations we can state that our estimates can be
performed in the Minkowskian tangent space. This is a 
consequence of the fact that
gravitons are travelling in a region whose size
is much smaller than that of the horizon.
Only for far clusters observed from distances
comparable to that of the horizon ($D_{0} \geq 600 \ Mpc $), 
the Minkowskian point of view 
is not valid and the approach used in this paper must be improved. 

Since our calculations can be carried out in the Minkowskian 
tangent space and galaxy clusters
are far of being relativistic in both senses,
special relativity, i.e., $v/c \le 10^{-3}$, and general relativity
$r_g/R \le 10^{-4}$ ($r_g =2GM/c^{2}$ is the Schwarzschild 
radius of the object), these clusters can be described by
using the so-called slow-motion formalism.  
The spacetime metric can be linearized in the usual way 
($g_{\mu \nu} = \eta_{\mu \nu} +
h_{\mu \nu}$), the transverse traceless (TT)
gauge can be used and, everywhere outside the cluster 
(Misner, Thorne \& Wheeler, 
1973), the spatial components of $h_{\mu \nu}^{^{TT}}$ are
\be
h_{ij}^{^{TT}}= \frac {G}{c^4}\frac {2}{D}\ddot{I}_{ij} (t-\frac{D}{c})
\label{hij}
\en 
where $I_{ij}$ are the components of the traceless inertial tensor. The 
contribution of the baryonic fluid, $I_{ij}^{^{B}}$, is 
$\displaystyle{I_{ij}^{^{B}}=
\int \rho x_i x_j d^3x - {1\over 3} \delta_{ij}\int \rho x^2 d^3x}$,
where $\delta_{ij}$ is Kronecker's delta, $\rho $ is the baryonic 
density, and $x$ the physical coordinates.
The CDM contribution, 
$I_{ij}^{^{DM}}$, is given by the following summation extended over all the
CDM particles,$\displaystyle{ 
I_{ij}^{^{DM}}= \sum_{n_p} m_p (x_i x_j - {1\over 3} \delta_{ij}  x^2)}$,
$n_p$ being the number of CDM particles and $m_p$ the mass of each particle. 
Finally, we compute
$I_{ij}=I_{ij}^{^{B}}+I_{ij}^{^{DM}}$.
Outside the cluster, the relative motion of two neighbouring particles A
and B moving with the cosmological background (ideal detector) is fully 
described by the quantities $h^{^{TT}}_{ij}$. 
In TT gauge, there is a system of coordinates
attached to A, in which the coordinate variations of the particle B are
\be
X^{i}_{_{B}}(t)-X^{i}_{_{B0}} = \frac {1} {2} X^{j}_{_{B0}}
[h^{^{TT}}_{ij}(t)]_{_{A}}
\label{detec}
\en
where $X^{i}_{_{B0}}$ stands for the initial coordinates of the particle
$B$ and the quantities $h^{^{TT}}_{ij}$ 
are calculated at point $A$. From this formula, 
it follows that oscillations in the $h^{^{TT}}_{ij}$
quantities lead to oscillations in the relative position of particles A and 
B with related frequencies and amplitudes.

In the slow motion approximation, the gravitational luminosity, $L_{GW}$, is 
given by the well known formula
\be
\label{lum}
L_{_{GW}}={G\over {5c^{5}}} \sum_{ij}^{1,3} 
{\stackrel{\displaystyle{...}}{I}}_{ij}
{\stackrel{\displaystyle{...}}{I}}_{ij}  \ .
\en
Since $I_{ij}$ is the addition of two terms, the
gravitational luminosity can be decomposed in the following evident manner, 
$\displaystyle{
L_{_{GW}}=L_{_{GW}}^{^{B}}+L_{_{GW}}^{^{DM}}+L_{_{GW}}^{^{CROS}}}$.
A direct calculation of $L_{GW}$ based on Eq.(\ref{lum})
would be very problematic  
(Finn and Evans, 1990) as a result of the difficulties coming from the 
numerical noise inherent to the numerical computation of third order time 
derivatives. However, as several authors have suggested (see, e.g.,
M\"onchmeyer et al. 1991), the second order time derivatives involved 
in $\ddot{I}_{ij}$ can be written in terms of the quantities $\ddot{\rho}, 
\dot{\rho}, \ddot{x}$ and, $\dot{x}$, which, in its turn, can be expressed
in terms of related variables, taken advantage of the corresponding
system of equations governing the motion of baryonic and CDM components.

\section{Results}

Figure 1 shows our numerical estimates of $L^{^{DM}}_{_{GW}}$ (crosses),
$L_{_{GW}}^{^{B}}$ (triangles) and $L_{_{GW}}$ (piecewise continuous line) 
versus quantity $(1+z)^{-1}$. 
At redshifts between $30$ and $25$, 
CDM and baryons have roughly the same velocity profiles 
and proportional (according to their relative abundances) density fields. 
As a consequence, the mass is the only relevant parameter and
quantity $L_{_{GW}}^{^{B}}$ is smaller than $L_{_{GW}}^{^{DM}}$.
Since evolution leads to a CDM configuration more stationary and spherically
symmetric than the baryonic one,
the luminosity $L_{_{GW}}^{^{B}}$ becomes dominant at $z \sim 9$.
The total luminosity 
$L_{_{GW}}$ is a little greater than  $10^{36} \ erg/s$ in the redshift
intervals (25,30) and (3,7). 
The fact that luminosities appear to be similar in both periods 
can be easily understood taking into account that,
in the first interval, deviations from 
spherical symmetry are greater than in the second one, while dynamics 
is more violent between redshifts $7$ and $3$.
The luminosity, assumed constant, of a source radiating a total energy of 
$E_{_{GW}} \sim 0.15 \ M_{\odot} c^2$ during the age of the Universe
is $\sim 10^{36} \ erg/s$ . This means that
the efficiency of the gravitational-wave emission from the cluster 
is $\varepsilon = \displaystyle{\frac{E_{_{GW}}}{M c^2}}\sim 10^{-16}$.
Therefore, we cannot expect any energetically significant 
gravitational-wave background generated by galaxy cluster
evolution.

Quantities $h_{+}=h^{^{TT}}_{xx}$ and $h_{\times}=h^{^{TT}}_{xy}$ 
have been evaluated along the world line of an observer moving with the 
background who carries the ideal detector of Section 3. 
The z-axis has an arbitrary direction and the 
gravitational wave travels along this axis. Quantities $h_{+}$ and 
$h_{\times}$ oscillate with time having very large characteristic periods. 
Taking into account that, according to Eq. (\ref{hij}), 
quantities $h_{+}$ and $h_{\times}$ are proportional to $D^{-1} \propto 1+z$,
one easily concludes that the variations of amplitudes are proportional to 
$1+z$ (effect of background expansion). Hence, the oscillations of the 
quantities $h_{+}/(1+z)$ and $h_{\times}/(1+z)$ also have varying amplitudes,
but the variations of these new amplitudes 
only depend on the cluster evolution 
itself. These quantities are plotted, as a function of $t$, in Fig. 2. 
 
It is noticeable that the luminosities and amplitudes given in Figs. 1
and 2 are compatible with  rough estimates 
based on formulae from other astrophysical scenarios; in fact,
according to Shapiro \& Teukolsky (1983) the gravitational-wave
luminosity of a self-gravitating system of mass M, size R and typical
velocity $v$ is 
\be
L_{_{GW}} \approx L_{0} \displaystyle{\left(\frac{r_{g}}{R}\right)^2 
\left(\frac{v}{c}\right)^6}   \ ,
\label{LGW}
\en
\noindent
being $L_{0} = c^5/G \approx 3.63 \times 10^{59} erg/sec$.
For our simulated cluster is $\displaystyle{\left(
\frac{r_{g}}{R}\right) \approx 10^{-4},\,\,
\left(\frac{v}{c}\right) \approx 2 \times 10^{-3}}$, and Eq. (\ref{LGW}) 
leads to $L_{_{GW}} \approx 2\times 10^{35}\, erg/sec$. 
Moreover, the relative strain or amplitude of the 
gravitational wave signal
coming from a collapsing object at a distance $D$ of the terrestrial
detector -- in a form well-adapted to 
our cosmological scenario -- is (Shapiro \& Teukolsky, 1983)
\be
h \approx 3 \times 10^{-11} \displaystyle{(\frac{\varepsilon}{10^{-15}})^{2/7} 
(\frac{M}{10^{15} M_{\odot}}) (\frac{D}{100 Mpc})^{-1}}
\label{h2}       \ .
\en
\noindent 
For our simulated cluster ($\varepsilon \approx 10^{-16}$),
Eq. (\ref{h2}) leads to $h \approx 0.8 \times 10^{-11}$. Finally,
Schutz (1997) gives the following alternative formula:
\be
h \approx 5 \times 10^{-11} \displaystyle{
(\frac{E_{_{GW}}}{10 M_{\odot} c^2})^{1/2}
(\frac{f_{_{GW}}}{10^{-17} Hz})^{-1/2} (\frac{D}{150 Mpc})^{-1}}
\label{heff2}  \ ,
\en
\noindent
where $f_{_{GW}}$ is the frequency of the wave.
In the case of our cluster, the characteristic
dynamical time is $\approx t_0^{-1}$. This gives 
$f_{_{GW}} \approx 10^{-17} \ Hz$. Furthermore, 
we have $E_{_{GW}} \approx 0.15$, and $D = 100 Mpc$.
Taking into account all these data, Eq.
(\ref{heff2}) leads to 
$h \approx 10^{-11}$.

After describing Figs. 1 and 2 and analyzing their contents and
consistency, let us focus our attention on an important 
consequence.
The characteristic periods of the waves coming from the cluster are larger 
than $10^{9}$ years and, obviously, it is not possible to detect any 
complete pulse. However, we are going to show that the oscillatory fields 
$h_{+}$ and $h_{\times}$ produce an effect which could be observable -- during 
an admissible integration time -- with a standard detector like the one
described in Eq. (\ref{detec}). In fact, during the time interval $\Delta t$, 
the distance between two particles A and B aligned with the {\it x} or 
{\it y} axis 
undergoes a relative variation $\frac {1} {2} [dh_{+}/dt]_{0} \Delta t$ 
and, when the particles are aligned with the directions forming angles of 
$45^{\circ}$ with these axis, their relative distance have changed in a
quantity given by $\frac {1} {2} [dh_{\times}/dt]_{0} \Delta t$. The
above derivatives -- to be performed at present time -- can be estimated from 
the data displayed in the curves of Fig. 2. The resulting values are
$[dh_{+}/dt]_{0} = -1.26 \times 10^{-20} \ yr^{-1}$ and 
$[dh_{\times}/dt]_{0} =  -4.65 \times 10^{-21} \ yr^{-1}$. In brief,
our simulated cluster produces changes in the relative distance of the order 
of $10^{-22}$ -- detectable with current technology -- in a short period of 
{\em four days}. This variation in the relative distance is an effect 
which would last -- at the same rate -- for many years; in this sense,
we can speak about a {\em secular effect}. The relative
separation distance would vary 
regularly reaching the order of $10^{-19}$ after one decade. 

\section{Conclusions and Discussion}

We have proved that gravitational radiation from clusters does not
contribute significantly to the density parameter. 

A new effect produced by the gravitational waves
emitted by a cluster has been described. 
A cluster with the features of the
observed ones has been simulated in the standard CDM model.
In other cosmogonies,
the effect produced by clusters having these features 
is expected to be comparable. This is because
the clusters producing the secular effect are located
near the observer and,
consequently, their evolution --well inside the horizon-- is dominated 
by internal interactions.
Simulations support this idea, see, e.g., Huss, Jain \& Steinmetz (1998), 
who claim that the
differences between individual cluster realizations 
of a given  model are 
more important than the differences within differing scenarios
for a given realization. However, 
the properties of the spatial distribution of clusters depend on the chosen 
scenario strongly. 
A few words about 
detection and future work are worthwhile.

We could take advantage of the fact that the secular effect 
can be observed for 
many years. In order to do that, 
this effect should be measured during periods 
greater than the characteristic time of any 
time varying gravitational field acting on the detector. 
Thus, the pulses produced by these fields could be seen as local 
perturbations of the secular effect.

If the detector is pointing towards a certain direction, it is not
receiving gravitational waves from an unique cluster; hence,
the calculation of the 
total effect produced by a set of clusters is
important. For the sake of briefness,
we present a rough estimate based on
idealized cluster distributions within a
sphere of radius $R_{s}$. All the clusters are identical to that
of this paper. The distance between any pair of clusters is constrained to 
be larger than a fixed distance 
$L$. No cluster correlations are considered.
Given an observation direction, the contribution of
each cluster to the
relative variation of the distance between particles $A$ and $B$
(detector)
is assumed to be $\frac {1}{2} \Psi \Delta t$ with
$\Psi = 1.26 \times 10^{-20}\, (\frac {100}{D})\,
\kappa \, cos \alpha \,\, yr^{-1}$,
where $\alpha$ is the angle 
formed by the line of sight of the cluster and the observation
direction. 
Since clusters do not radiate
in a coherent way, their contributions to $\Psi$ are not correlated.
This fact is simulated by using the
random number $\kappa$ uniformly distributed in the interval (-1,1).
The factor (100/D) has been introduced because 
the effect of a given cluster
decreases with distance as $D^{-1}$ (see Eq. (\ref {hij})). 
After superimposing the contribution of all the clusters,
the mean $\mathcal{M}$
and the standard deviation $\sigma$ have been calculated from 
the $\Psi$ values 
corresponding to many observation directions.
For each pair ($R_{s}$, $L$), various simulations have been done.
The mean $\mathcal{M}$ changes from one 
to another, but $\sigma$ is a very stable quantity.  
For the pairs 
($R_{s}=600 \ Mpc$, $L=50 \ Mpc$),
($R_{s}=600 \ Mpc$, $L=100 \ Mpc$),
($R_{s}=500 \ Mpc$, $L=50 \ Mpc$),
we have found $\sigma=7.6 \times 10^{-20}$,
$\sigma=2.7 \times 10^{-20}$ and, 
$\sigma=7.0 \times 10^{-20}$, respectively. From these data
one concludes that, for $R_{s}\ge500$,
quantity $\sigma$ depends on $R_{s}$ weakly. This fact suggests that
the use of radius $R_{s}>600$ is not necessary because
the main part of the effect is produced by nearby clusters.
Furthermore, the resulting 
$\sigma$ values are greater than the value $1.6 \times 10^{-20}$
--maximum value assumed for a single cluster and D=100-- even when 
large separations
($L=100 \ Mpc$) are assumed. The large $\sigma$ values
given by simulations
strongly  suggests the feasibility of anisotropy measurements.
Improved simulations including 
cluster correlations should increase the anisotropy. 
The effect of realistic 
spatial distributions 
including correlations and clusters with different masses, 
sizes, et cetera will be studied elsewhere. 
Various parameters --as the 
proportions between baryonic and dark matter, the density parameter,
et cetera-- would be involved in these simulations. 
The resulting anisotropy would 
depend on both cosmological and large scale structure parameters.
Comparisons between measurements and predictions of the secular effect
and its anisotropies could play a crucial role 
in Modern Cosmology.

\acknowledgements 
This work has been 
supported by the Spanish DGES (grants PB96-0797 and PB94-0973).
V. Quilis thanks to the Conselleria d'Educaci\'o i 
Ci\`encia de la Generalitat Valenciana for a fellowship.
Calculations were carried out in a SGI O2000 
at the {\it Centre de Inform\'atica de
la Universitat de Val\`encia}.

\newpage
 
 
\figcaption{Gravitational-wave luminosity as a function of redshift $z$.
Piecewise continuous line , crosses and triangles correspond to 
the total ($L_{_{GW}}$), dark matter ($L_{_{GW}}^{^{DM}}$),
and baryonic ($L_{_{GW}}^{^{B}}$) luminosities,
respectively.}

\figcaption{Quantities $h_{+}/(1+z)$ and $h_{\times}(1+z)$ 
as functions of the cosmological time $t$.}

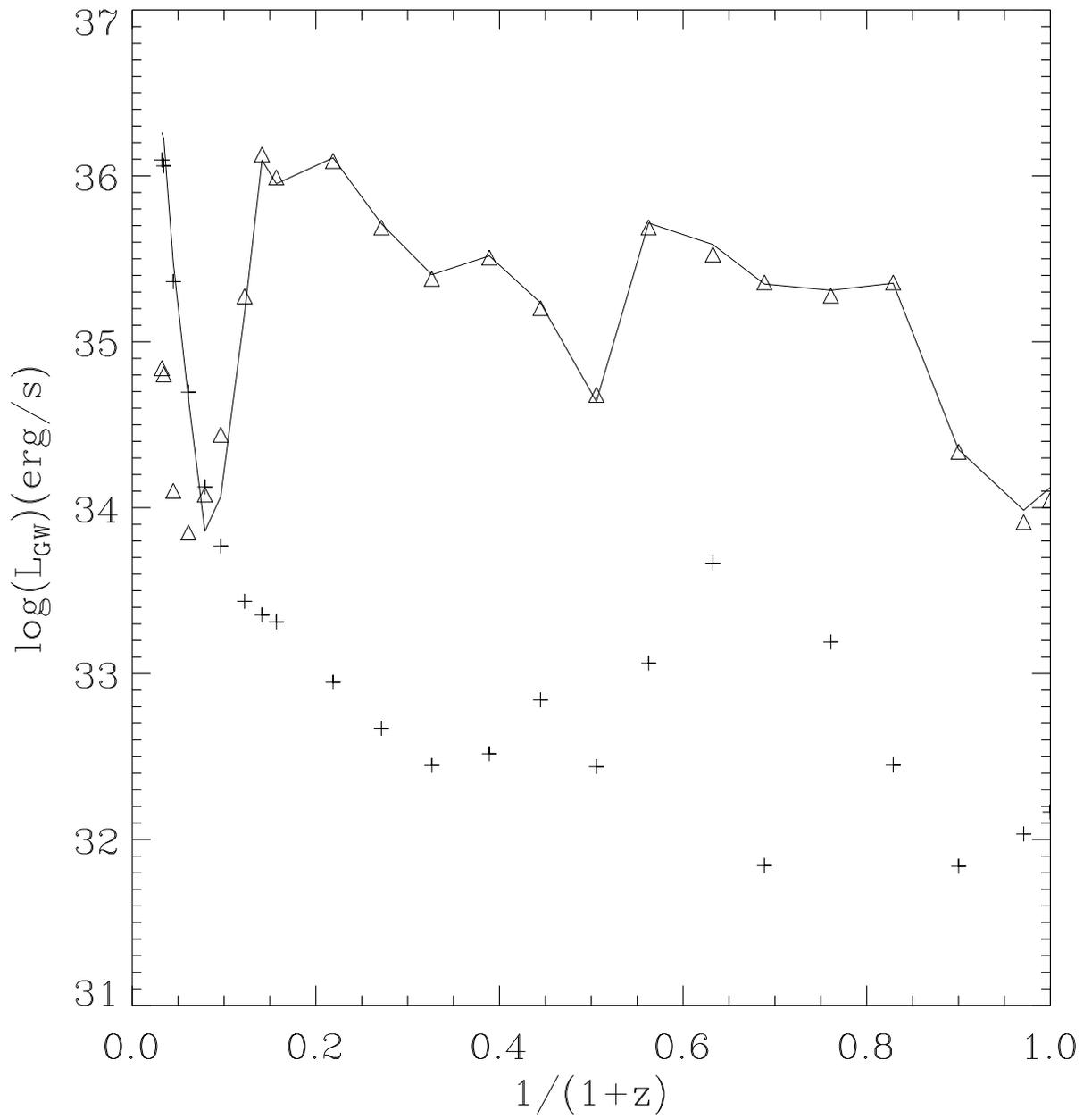
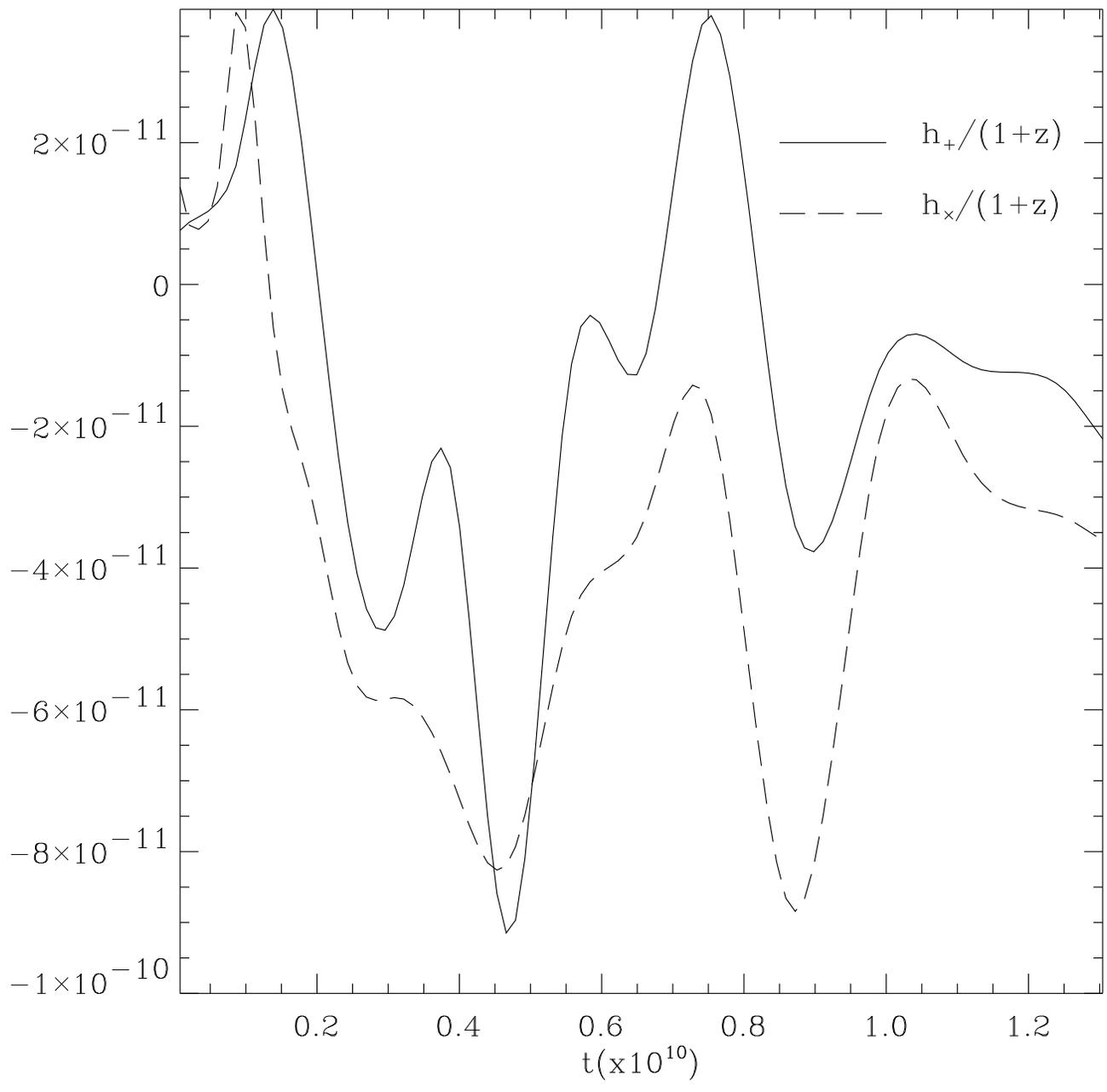
\end{document}